\PassOptionsToPackage{utf8}{inputenc}
\documentclass{article}
\usepackage{authblk}
\usepackage{indentfirst} 
\usepackage{url}
\usepackage{amsthm}
\usepackage{textcomp}
\usepackage{tabularx}
\usepackage{verbatim}
\usepackage{graphicx}
\usepackage{soul}
\usepackage{amsmath}
\usepackage{amssymb}
\usepackage{mathabx}
\usepackage{enumitem}
\usepackage {tikz}
\usepackage{tikz-uml}
\tikzstyle{rect} = [rectangle, draw, fill=white!20, node distance=2.5cm, text width=6em, text centered, sharp corners, minimum height=3em]
\usetikzlibrary{plotmarks,intersections,calc,decorations.pathreplacing,positioning,arrows,shapes.geometric,matrix,decorations.markings}
\usetikzlibrary {positioning}
\newcommand{\pubmed}{\text{PubMed}}
\newcommand{\pubmedR}{$\mathbb{PM}$}
\newcommand{\pmid}{\textsc{pmid}}
\newcommand{\doi}{\textsc{doi}}
\newcommand{\neurSyn}{$\mathbb{SS}_{ns}$}
\newcommand{\neuroboun}{\textit{NeuroBoun}}

\DeclareMathSymbol{\mathdblquotechar}{\mathalpha}{letters}{`"}

\newcommand{\mathdblquote}{\mathtt{\mathdblquotechar}}
\begingroup\lccode`~=`"\lowercase{\endgroup
  \let~\mathdblquote
}

\def\kbbox[#1,#2,#3,#4,#5]#6{
        \draw[solid] node[draw,color=gray!50,minimum
        height=#1,minimum width=#2] (#4) at #5 {}; 
        \node[anchor=#3,above,inner sep=2pt] at (#4.#3)  {#6};
        }

\newcommand{\ie}{i.e,}

\newcommand{\nt}[1]{\texttt{#1}}
\newcommand{\fn}[1]{\textnormal{#1}}

\newcommand{\queryt}[1]{\texttt{"}\textit{#1}\texttt{"}}
\newcommand{\timeint}[2]{\ensuremath{\left<#1,#2\right>}}

\newcommand{\inq}[5]{$\langle$#1,$\{$ #2 $\}$, #3, #4, #5$\rangle$}

\newcommand{\inqvar}[5]{\ensuremath{\langle #1, #2 , #3, #4, #5 \rangle}}

\theoremstyle{plain}

\newcommand{\var}[1]{{\scriptsize \ensuremath{\mathit{#1}}}}
\newcommand{\qvar}[1]{{\scriptsize \ensuremath{\mathit{#1}}}}
\newcommand{\type}[1]{\nt{#1}}
\newcommand{\vart}[2]{\var{#1}$\colon$\type{#2}}
\newcommand{\setvar}[1]{\type{#1}\ensuremath{_s}}
\newcommand{\starset}[1]{\var{#1}\ensuremath{\ast}}
\newcommand{\plusset}[1]{\var{#1}\texttt{+}}

\theoremstyle{definition}

\newcommand{\fmri}{fMRI}
\newcommand{\mesh}{MeSH}

\begin{document}
\title{NeuroBoun: An inquiry-based approach for exploring scientific literature --\\ a use case in neuroscience}
\date{}

\author[$\dagger$]{Uskudarli, S\thanks{Corresponding author, suzan.uskudarli@boun.edu.tr}}
\author[$\star$]{G{\"o}kdeniz, E.}
\author[$\ddag$]{Canbeyli, R.}



\affil[$\dagger$]{Department of Computer Engineering, Bogazici University, Turkey}
\affil[$\star$]{Sony Global Solutions, Turkey}
\affil[$\ddag$]{Department of Psychology, Bogazici University, Turkey}

\maketitle

\abstract{Online scientific publications provide vast opportunities for researchers. 
Alas, the quantity and the rate of increase in the articles make the utilization of these resources very challenging. 
This work presents as inquiry-based approach to support the articulation of complex inter-related queries to gain insights regarding how these subjects have been studied in conjunction with one another as reported in the scientific literature.
For this purpose we introduce inquiries that represent  inter-related subqueries that are of interest to a researcher.
The inquiries are expanded to better capture the intent of the inquirer, from which several queries are generated that represent various juxtapositions of the subjects in consideration.
The sets of queries are used to search repositories to yield results that reveal quantitative and temporal relations among the subjects of the inquiry.
A web-based tool, \neuroboun,  is developed as a proof of concept for medical publications found in \pubmed. A use case related to the asymmetry of amygdala is presented to illustrate the potentials of the proposed approach. 
}

\section{Introduction}
\label{sec:introduction}

Gaining insight into the nature and results of a scientific investigation is quite complex. 
Additionally, the accelerating rate of scientific publications make it exceedingly difficult to follow scientific developments.
To grasp the developments related to a particular subject, an inquirer needs to be able to glean the relevant aspects and results that have been contributed by numerous scientists over a period of time.
For example, complex and comprehensive search (inquiries) consider the case of amygdalar asymmetry in the presence of themes that are often of interest, such as mood (i.e. anxiety, depression, etc.), brain functions (learning, memory, etc) and evaluative aspects (salience, valence, etc).
The process of inquiry requires support for accessing information at various resolutions.
To this end, one may need to first explore various concepts related to the inquiry (high level) and then delve deeper into details (low level). Moreover, the inquiry process must be responsive so as to support the flow of the inquiry, since the responses will likely guide further inquiries.

This work originates from a real-world use case in the neuroscience domain where we investigate the lateralization of brain regions as to how extensively they have been so far studied. 
We had been systematically querying \pubmed\ to understand how certain concepts in neuroscience are inter-related and to what degree they have been investigated by examining publications. 
This process was laborious, time consuming and potentially error prone since such exploration is performed via aggregating the results of numerous manual queries. 
Upon reflection, it was evident that a tool that supports the articulation of interrelated queries that would automatically query \pubmed\ could relieve all three concerns and expedite the topical search without disrupting the flow of the researcher’s attention. 
The design of \neuroboun\ emerged while investigating the cerebral asymmetry in general and the lateralization of subcortical structures such as the amygdala in particular. 

To address such needs we introduce the notion of an \textit{inquiry} to express a set of inter-related queries with a specific context that is of interest to a researcher.
Inquiries are made within a collection of documents that represent a body of work (documented knowledge) and may be domain dependent (i.e. medical publications) or  independent (i.e. news articles).
The objective of an inquiry is to provide insight regarding the research performed on inter-related subjects over time, which requires posing several advanced searches. 
Essentially, inquiries consist of a main query which serves as the central focus of interest (i.e., the brain region \textit{amygdala}) and an optional set of relevant queries (i.e., the imaging technology \textit{fMRI} and the mood disorders \textit{depression and anxiety}).
The inquires are processed to better capture the intention to yield a set of queries, the results are which are aggregated to present an explorable overview to the inquirer. 

For this purpose, we created a tool called \neuroboun\ tailored to address the need to run inquiries for researchers to perform several comparisons without losing the context of their search.
A prototype is implemented to demonstrate this approach using \pubmed~\cite{pubmed-help} as the source of documents, which consists of over 29 million articles related to biomedical literature as of July 2019. 
This tool was not designed as an alternative to \pubmed, but rather to enhance and expedite the searching of topics within the body of work it contains.

The need for complex and comprehensive search (inquiries) is illustrated using \neuroboun\ with a use case concerning \textit{amygdalar asymmetry} in conjunction with relevant themes of mood (anxiety, depression), brain functions (learning, memory, etc) and evaluative aspects (salience, valence, etc).

The remainder of this paper is organized as follows: Section~\ref{sec:relatedwork} introduces related work, Section~\ref{sec:approach} presents our approach, Section~\ref{sec:implementation} outlines the implementation details, Section~\ref{sec:use_case} elaborates our use case,  Section~\ref{sec:discussion} presents our observations related to this work and the future directions we intend to pursue, and finally  Section~\ref{sec:conclusions} makes concluding remarks.

\section{Related Work}
\label{sec:relatedwork}

There is great motivation to improve access to research results, which has led to work on various aspects of this vast challenge.
This section describes some of this work in terms of their domain of subjects, the methods for performing searches, and the presentation of the results.

Systems that support access to research publications may be specific to or independent of a domain.
Google Scholar~\cite{google-scholar} is a widely used domain-independent service that supports querying scholarly work (journal and conference papers, theses and dissertations, academic books, pre-prints, abstracts, technical reports) and patents. 
It returns a ranked list of articles with the number of citations they received and related work. 
Semantic Scholar~\cite{semantic-scholar-website} is also a domain-independent service that utilizes artificial intelligence (AI) an provides functionality similar to Google Scholar. 
It provides intent-based searching that uses features like locations, variation of words, synonyms and concept matching to yield relevant peer reviewed articles.
Google Scholar has higher recall and lower precision in comparison to Semantic Scholar due to its coverage and  search methodology. 
Users elect to use the one based on the scope and depth of their intended search.
Sometimes Semantic Scholar is used to leap frog to Google Scholar to fetch additional articles.

Among the most actively utilized scientific documents are those from the biomedical domain due to the immense advances that have taken place  in this field coupled with the desire to make timely use of these developments in life impacting medical research.
\pubmed (\cite{pubmed-help}) provides a continuously evolving platform for exploring  scholarly documents related to biomedicine and health (presently over 30 million articles) with sophisticated search functionality.
To capture the intended search,  the query is enriched with alternative terms obtained from translation tables such as MeSH (Medical Subject Headings) and other indexes such as for authors. 
\pubmed\ also provides  application programming interfaces (APIs) called the \textit{Entrez Programming Utilities API} \cite{Entrez} for the developers interested in utilizing \pubmed\ data  and querying the \textit{National Center for Biotechnology Information} (NCBI) databases. 
Wrapper applications using the \textit{Entrez API} \cite{fontelo2005AskMedline, Muin2006PubMedIA} aim to improve the queries, the relevancy of search results, or the user experience. 

Several alternatives to using \pubmed\ for searching \textsc{medline} \cite{toolComparisonKeepanasseril,toolComparisonLu}  were found to improve the comprehensibility of the results by novice or expert users via the providing filters and visualizations. 
Most of them return lists of ranked articles with various attributes.
A few return tables or graphs that relate predetermined features such as domain specific concepts (proteins or genes), authors, \mesh\ terms, or locations.
Our work differs from these approaches as our focus aims to support user defined searches with inter-related dimensions and to respond with an overview that depicts how these dimensions are related. If desires, the researcher can continue to diving in to specific aspects of result to eventually retrieve articles.

Most of the recent work, including new and improved services, focus on smart search technologies tailored to individual researchers. 
Artificial intelligence (AI) techniques are employed to deliver precise, comprehensive and useful output.
As the main source documents in the biomedical domain,  \pubmed\ \cite{pubmed20}  is continuously being improved in preparation for its next generation model. 
\pubmed\ Labs \cite{pubMedLabs} is the test site for the new/improved functionalities such as ranking algorithms and optimized user-interface experience. 
Meta \cite{meta} is developing a service to assist researches with customized feeds that deliver real-time developments (publications, researchers, and related concepts) in biomedical research, thereby reducing the laborious tasks of tracking research. 
Similarly, Scopus Discovery \cite{scopus-discovery} is developing a model to provide personalized content discovery that is refined through learning researchers' behavior based on their previous activities.

All these platforms focus on one of the most important aspects of scientific literature search,  namely identifying the intent of a researcher by examining their behaviour patterns. Understanding the intended query is crucial to delivering desired results. 
Prior to the recent AI-based improvements log analysis \cite{logAnalysisHerskovic, islamaj2009logAnalysis} or session based approaches \cite{Levine:2017:ERM:3077136.3080664,VanGysel:2016:LQM:2970398.2970422} were employed to understand searching behavior patterns.
Such behaviour was found to differ based on the domain of research, such as in oncology \cite{Vincent2006MakingPS} and pandemic diseases \cite{Norgaard2010SearchingPD}. 
. 


Some studies focus on extracting relations and connectivity information for specific domains to automate the retrieval of precise and comprehensive information relevant to a query performed against an extensive body of scientific information. 
WhiteText \cite{whitetext-web} extracts neuroanatomical connectivity relations from publications retrieved from \pubmed~\cite{french2015text} using  NLP techniques.
It presents the connected regions and species as a result.
To preserve the context, the Named Entities (proteins, genes, and brain regions) are extracted from the documents on a sentence basis.  
A similar approach extracts neuroanatomical relations by specifying patterns over the constituency and dependency parse trees of sentences~\cite{gokdeniz2016automated}. 

\neuroboun\ is a wrapper application that uses \pubmed\ as a data source that aims to enhance the existing capabilities of \pubmed\ and expedite inquiries which are complex search operations related to inter-related subjects. 
The main difference of our approach is the generation of a set of subqueries from inquiries and the aggregation of the result to provide an overview of how the subqueries are inter-related. The overview can be further explored to see the details and eventually reach the articles that contribute to the results. 
This overview may be used to focus on particular sub-sets or to revise the inquiry to alter the results. 
\neuroboun\ provides a comprehensive and contextual user journey for the researchers as is detailed with a use case in Section~\ref{sec:use_case}. 
The ultimate goal is to provide researchers with a tool that enables them to concentrate on their investigation without having to switch contexts.

\section{Approach}
\label{sec:approach}

We introduce the notion of an \textit{inquiry} to express a set of inter-related queries with a specific context that is of interest to a researcher.
Inquiries are made within a collection of documents that represent a body of work (documented knowledge) and may be domain dependent (i.e. medical publications) or  independent (i.e. news articles).
Essentially, inquires have a main query which serves as the central focus of interest, such as the brain region \textit{amygdala}.
Additionally,  other queries of interest in conjunction with the main query may be specified. 
For example, queries related to the \textit{amygdala} may include the imaging technology \textit{\fmri} and the mood disorders \textit{depression and anxiety}.
These queries could be of interest when studying the relevance of imaging technologies and mood disorders in research related to amygdala.

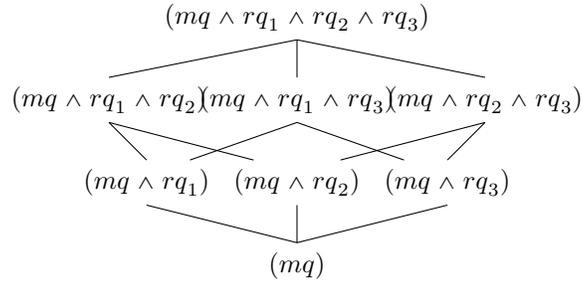
\begin{figure}
\begin{center}
\label{fig:queries-of-inquiry}
\begin{tikzpicture}
    \node (everything) at (0,0) {$(mq\wedge {rq}_{1} \wedge   {rq}_{2}   \wedge {rq}_{3})$};
    
    
        \node [below=0.5cm  of everything] (mqf1f3) {$(mq \wedge {rq}_{1} \wedge  {rq}_3)$};  
        \node [left of=mqf1f3,xshift=-1.5cm] (mqf1f2)  {$(mq \wedge  {rq}_{1} \wedge {rq}_2)$};
        \node [right of=mqf1f3,xshift=1.5cm] (mqf2f3) {$(mq \wedge {rq}_{2} \wedge  {rq}_3)$};

    
        \node [below=0.5cm of mqf1f3] (mqf2) {$(mq\wedge {rq}_{2})$};
        \node [left  of=mqf2,xshift=-1cm] (mqf1)  {$(mq\wedge  {rq}_1)$};
        \node [right of=mqf2,xshift=1cm] (mqf3) {$(mq\wedge {rq}_{3})$};

  
    \node [below=0.5cm of mqf2] (bottom) {$(mq)$};
    

     \draw [] (everything.south) -- (mqf1f2.north);
     \draw [] (everything.south) -- (mqf1f3.north); 
     \draw [] (everything.south) -- (mqf2f3.north);     
    
    \draw [] (mqf1f2.south) -- (mqf1.north);
    
    (c1)!0.5!(c2)
    \draw [] (mqf1f2.south) -- ($(mqf2.north west)!0.2!(mqf2.north east)$);
    \draw [] (mqf2f3.south) -- ($(mqf2.north west)!0.8!(mqf2.north east)$);
    \draw [] (mqf1f3.south) -- ($(mqf1.north west)!0.8!(mqf1.north east)$);
    \draw [] (mqf1f3.south) -- ($(mqf3.north west)!0.2!(mqf3.north east)$);
    \draw [] (mqf2f3.south) -- (mqf3.north);

    \draw [] (mqf1.south) -- (bottom.north);
    \draw [] (mqf2.south) -- (bottom.north);
    \draw [] (mqf3.south) -- (bottom.north);
     
\end{tikzpicture}
\caption{The queries of  an inquiry with three aspects (${rq}_1, {rq}_2, s_3$) and the main query $mq$.}
\end{center}
\end{figure}

The objective of an inquiry is to provide insight regarding how various queries are related to one another over time.
This involves examining the queries in conjunction with one another. 
Since the main query establishes the context of the inquiry, it is present in all query result.
The remaining queries are performed in all possible combination with each other.
Figure~\ref{fig:queries-of-inquiry} shows  the queries related the inquiry \inq{mq}{$rq_{1},rq_{2},rq_{1}$}{c}{s}{ti}, where $mq$ and $rq_{i}$
represents the main query and related queries respectively.

\begin{figure*}[t]

    \newcommand{\OverviewScale}{0.7}

    \begin{center}

        \tikzstyle{decision} = [diamond, draw, fill=white!20, text width=4.5em, text badly centered, node distance=3cm, inner sep=0pt]
        \tikzstyle{block} = [rectangle, draw, fill=white!20, node distance=2.6cm,text width=4.9em, text centered, rounded corners, minimum height=4em]
        \tikzstyle{line} = [draw, -latex']
        \tikzstyle{cloud} = [draw, ellipse,draw, fill=white!20, node distance=2cm,
                text width=6em, text centered, rounded corners, minimum height=4em]
        \tikzstyle{database} = [cylinder,cylinder uses custom fill,cylinder body fill=white!50,cylinder end fill=white!50,shape border rotate=90,aspect=0.25,draw]
        \tikzstyle{enclosure} = [rectangle, draw, dash dot, fill=white!20, node distance=3cm,minimum height=4em, align=center,text width=4.8em]   
                    
        \begin{tikzpicture}[node distance = 2cm, auto]
                       
        \umlactor[x=0,y=0] {user};
        \node [block, right of= user,yshift=1cm, xshift=2.6cm] (constructqueries) {Cross Queries};
        \node [block, right of= user,yshift=-1cm, xshift=2.6cm] (expandqueries) {Expand Queries};
                                    
        \coordinate (midCQ) at ($(constructqueries.north east)!0.5!(expandqueries.south east)$);
                                            
        \node [block, right of= constructqueries, node distance=2.8cm] (searchqueries) {Search Queries};

        \node [right of= searchqueries,node distance=2.8cm,yshift=-0.25cm] (r2) {$\var{q}_2\!:\!\{d_{21}, \hdots \}$};
        
        \node [above of= r2,node distance=0.5cm] (r1) {$\var{q}_1\!:\!\{d_{11}, \hdots \}$};
        
        \node [above of= r1,node distance=0.9cm, align=left] (results) {Result set:\\[-3pt] $\forall d \in $ \starset{d}};
        \node [below of= r2,node distance=0.5cm] (r3) {$\vdots$};

        \node [enclosure,  below of= user, xshift=1.7cm, node distance=1cm] (synonyms) {Synonyms\\Acronyms};
                                                                
        \node [enclosure , below of= synonyms, node distance=1.8cm] (docrepo) {Document Repository};
        \node [block, below of= expandqueries,node distance=1.8cm] (processdocs) {Process Documents};
        
        \node [database, below of= searchqueries, node distance=3.8cm, text width=4.8em] (docdb) {Documents};
        
        \path [->,>=angle 90]  (constructqueries) edge node[auto]{\var{cq}}  (expandqueries);
                                                                 
        \path [->,>=angle 90]  ($(expandqueries.west)-(0,0.2)$) edge node[above,xshift=-0.1]{{\scriptsize $q_s$}}  ($(synonyms.east)-(0,0.2)$);
        
        \path [->,>=angle 90]  ($(synonyms.east)-(0,0.6)$) edge node[above,xshift=-0.2]{\starset{s}}  ($(expandqueries.west)-(0,0.6)$);
                                                                                         
        \path [line] (searchqueries.240) -- node[left] {\var{q}}(docdb.120);
        
        \path [line] (docdb.62) -- node[right] {\starset{d}} (searchqueries.300); 

        \path [->,>=angle 90]  (searchqueries) edge node[auto]{} ($(r2.west)-(0.2,-0.2)$);
        
        \coordinate (topmid) at ($(searchqueries)+(0,1.9)$);
        
        \draw [->,dotted] (results.north) |- (topmid) -| (user.north);

        \path[->,>=angle 90] (docrepo)  edge node[auto] {\starset{r}}  (processdocs) ;
        
        \path[->,>=angle 90] (processdocs) edge node[above,near start] {\starset{d}} ($(docdb.west)-(0,0)$) ;
 
         \draw[black,thick,dotted] ($(constructqueries.north west)+(-1,0.2)$)  rectangle    ($(expandqueries.south east)+(0.2,-0.2)$); 
        
        \node[above of= constructqueries,node distance=1.2cm]{Construct Queries};
    
         \draw[black,thick,dotted] ($(searchqueries.north west)+(-0.3,0.2)$)  rectangle    ($(docdb.south east)+(0.8,-0.2)$); 
         \node[above of= searchqueries,node distance=1.3cm,align=left]{Search\\[-3pt] $\forall q \in eq$};
         \draw[black,thick,dashed] ($(r1.north west)+(-0.05,0.2)$)  rectangle  ($(r3.south east)+(1,-0.2)$); 
         \draw[black,thick] ($(constructqueries.north west)+(-3.6,1)$)  rectangle    ($(docdb.south east)+(1,-0.5)$); 
         \coordinate (anchorCQI) at ($(user)+(3.2,0)$); \coordinate (anchorCRQI) at ($(constructqueries.west)-(0.4,0)$);
         \coordinate (anchorEQI) at ($(expandqueries.west)-(0.4,-0.3)$);
         \coordinate (anchorEQR) at ($(expandqueries.east)+(0.6,0)$);
         \path[-,>=angle 90] (user)  edge node[auto] {\var{i}}  (anchorCQI) ;
         \draw [->] (anchorCQI)  |-  (anchorCRQI) -- node[xshift=-0.2cm,align=left]{\qvar{i.mq}\\[-3pt]\qvar{i.rq}} (constructqueries.west);
         \draw [->] (anchorCQI)  |-  (anchorEQI) -- node[xshift=-0.2cm,align=left]{\qvar{i.syn}} ($(expandqueries.west)+(0,0.3)$);
         \draw [->] (expandqueries.east) -- (anchorEQR) |- node[right,near start]{\var{eq}} ($(midCQ)+(0.8,0.7)$) --  ($(searchqueries.west)-(0,0.3)$) ; 
\end{tikzpicture}
\caption{An overview of processing inquires. Here $i$ is an inquiry posed by a user, $cq$ is a set of queries composed by juxtaposing the related queries with the main query, and $eq$ is the corresponding set of queries expanded with synonyms.}
\label{fig:overview}
\end{center}
\end{figure*}
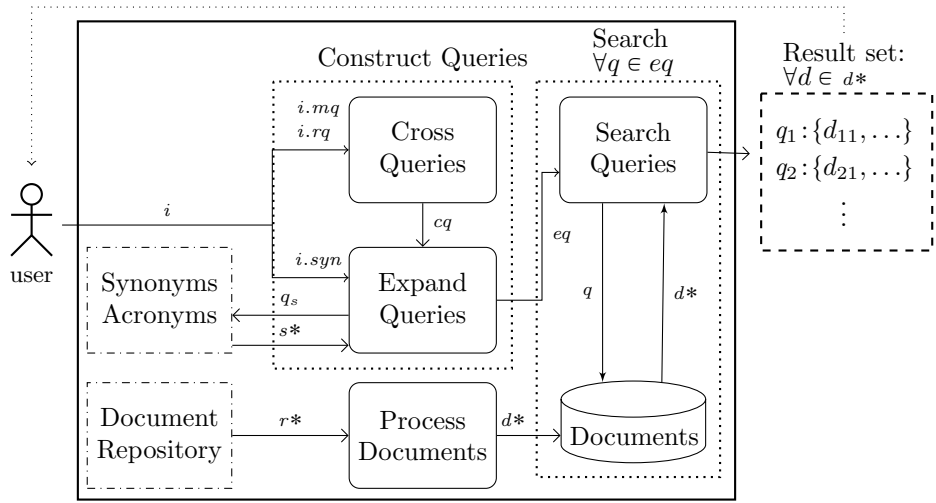

\begin{table}
\begin{center}
\begin{tabular}{lcl}
\hline
Type  &  Definition & Description\\ \hline
\nt{I} & $< \nt{MQ}, \nt{RQ}, \nt{DC}, \nt{SY}, \nt{TI} >$ &  inquiry\\
\nt{MQ} & \nt{QT}  &  main query \\ 
\nt{QT}& \nt{W}+ & term \\ 
\nt{RQ}& $\{ \nt{QT} \texttt{ ","} \}\star$  &  related query\\
\nt{W} & \textrm{w-regex} &  word \\ 
\nt{S} &  \textrm{s-regex} &  sentence  \\ 
\nt{TI} & $ <\nt{Y}, \nt{Y}>$ &  time period (interval)\\ 
\nt{Y}  &  \textrm{y-regex} & year. \\ 
\nt{R}&    \nt{D}$+$ &  document resource \\
\nt{D} & $<\nt{T},\nt{S\texttt{+}},\nt{Y},\nt{ID}>$ & document  \\ 
\nt{DC} & $\{ \nt{D} \texttt{ ","} \}\texttt{+}$  & document collection \\ 
\nt{T} & \texttt{String} & title \\
\nt{SYN}  & $\nt{SI}\star$ & set of synonyms items \\ 
\nt{SI}  & $<\nt{W},[\nt{W}+]>$ & synonyms \& acronyms  \\ \hline  
\end{tabular}
\caption{\label{tab:types} The type symbols used for making inquiries.}
\end{center}
\end{table}

Related queries may include sub-queries, each of which is searched independently.
For example, the inquiry \inq{mq}{$a_1,a_2$}{\texttt{D}}{\texttt{Syn}}{ti}, if ${rq}_1 = {rq}_{11}, {rq}_{12}$  and ${rq}_2={rq}_{21}$ the set of queries to be performed is: $\{ mq,mq \wedge {rq}_{11}, mq \wedge {rq}_{12}, mq \wedge {rq}_{21} \}$. 
Here, the sub-queries are not examined in conjunction with one another  ($ q_{11} \wedge  q_{12} $), but rather the dimensions are examined in conjunction with one another. 
The rationale here is that related queries capture an aspect of interest within the context of inquiry.
Thus, the aspects that are of interest with respect to each other are defined as separate dimensions rather than sub-queries.

The remainder of this section formally defines inquires by introducing  a set of types and functions.
Table~\ref{tab:types} shows the types relevant to inquiries.
All type names are denoted with capital letters.
Curly braces denote sets and angle brackets denote tuples.
The regular expression symbols ``\texttt{*}'' and ``\texttt{+}'' are used to denote sets of \textit{zero or more} or \textit{one or more} repetitions.

\subsection{Inquiries}

The main type \nt{I} (inquiry) consists of a main query (\nt{MQ}),  an optional set of related queries (\nt{RQ}),  a document resource (\nt{R}), a set of synonyms  (\nt{SYN}), and a time interval (\nt{TI}).
The default time interval corresponds to the dates of the oldest and  the more recent documents.
The document resource is a set of \nt{URI}s (Uniform Resource Identifier) from which information about publications are obtained.
Such information is the title, abstract, keywords, publication date, and the body of an article.
This paper describes inquires over the most widely accessible information about articles, namely their titles and abstracts.

Here, words, sentences and years are specified with regular expressions (s-regex, w-regex , and y-regex).  
A sentence consists of a sequence of characters that terminate with a ',', '!', or '?'\footnote{The prototype relies on the well known Natural Language Toolkit (\url{https://www.nltk.org/}) for sentences.} The regular expressions for words and years are defined as \verb|\W\\w+\W | and \verb|d{4}|.

A time interval is denoted as $<y_i,y_j>$ where  \vart{y_i}{Y}, \vart{y_j}{Y}, \var{y_i} $\leq$ \var{y_j}. 
A document \vart{d}{D} consists of a title, a set of sentences extracted from the abstract of the publication, a publication date, and an identifier provided by the document resource.
In the examples provided in this paper, the resource is \pubmed\ and the identifiers are called \textsc{PMID}\footnote{In our use case the resources are obtained from the Entrez API for \pubmed:  \url{https://eutils.ncbi.nlm.nih.gov/entrez/eutils/}}.

The semantics of performing an inquiry is defined with a set functions.
Variables are denoted with lowercase letters that are declared as \vart{variableName}{variableType} (i.e. \vart{inq}{I}).
Variables representing a set are denoted with the subscript ``s'' (i.e. \vart{\setvar{q}}{\plusset{QT}} for a set of query terms).
Accessor functions for composite data are presumed to exist and denoted using the dot notation with the name of the component. 
For example, $i.mq$ returns the main query of $i$ (\vart{i}{I}). 

In the following functions, let

\begin{equation*}
\mathrm{inquire}(i) = \inqvar{mq}{\setvar{rq}}{\setvar{d}}{\setvar{syn}}{ti}
\end{equation*}

\noindent where \vart{i}{I}, \vart{mq}{Q}, \vart{\setvar{rq}}{\plusset{Q}}, \vart{\setvar{syn}}{SY}, and \vart{ti}{TI}, 
\vart{\setvar{syn}}{SY} is a domains-specific set of synonyms and acronyms.

The main function $\mathrm{inquire}$ is defined as:
\begin{equation*}
\mathrm{inquire}(i) = \mathrm{search}(queries(i), \mathrm{documents}(i.dc,i.ti))
\end{equation*}

\noindent whose result set whose elements are the sets of documents that match a query of the inquiry. 
For example, the inquiry: 

$\mathrm{inquire}$(\inq{\queryt{amygdala}}
{\queryt{depression, anxiety},\queryt{\fmri}}
{\neurSyn}
{\pubmedR}
{\timeint{1980}{2018}})

\noindent \queryt{amygdala} (the main query) is examined along with the related queries: \queryt{depression, anxiety} and \queryt{\fmri}.
A set of synonyms (\neurSyn) is used to expand the query terms.
Also, the inquiry is limited to the documents published between $1980$ to $2018$ in accordance with:

\begin{equation}
\fn{documents}(d_{s},\timeint{b}{e}) =  \{ d | d \in d_{s}, b \leq d.y \leq e \}
\end{equation}

\vspace{0.5cm}

The query terms are expanded as:

\begin{equation}
\fn{expand}(qt,sy) =  {\displaystyle \prod_{i=1}^{|qt|} synset(w_{i},sy)} 
\end{equation}

 \noindent where $w_{i}$ is the $i^{th}$ word in $qt$ and $\mathrm{synset}(w,s)$ is the set of synonyms and acronyms for $w$ in $s$.








The queries corresponding to an inquiry $i$ are obtained by taking the cross product of the main query with the power set of the related queries (both expanded):

\begin{equation}
\fn{queries}(i) = \fn{expand}(i.mq) \bigtimes \mathcal{P}(\fn{expand}(i.rq,i.syn))
\end{equation}

\noindent where $\fn{queries}(i)$ forms the set of all queries that will be performed. We refer to an element of this set as a \textit{cross-query}. 
The search function for a query term is:

\begin{equation}
\fn{search}(q_s,d_s) = \{ d | d \in d_s, q \in q_s, match(q,d)=true \}
\end{equation}

\begin{equation}
    match(q,d,\Delta w) = \textrm{s-match}(q,d.tit,\Delta w) \lor  \textrm{s-match}(q,s_i,\Delta w)
\end{equation}

\noindent where , $s_i \in \fn{d.sentences}$ and  $\fn{s-match}$ is a function that returns \text{true} if a query term is contained in a string and \textit{false} otherwise. 
For query terms of single words, this is equivalent to the word being an element of the set of words within the document. 
For multi-word terms, the words must occur in the document in the same order as in the query. 
Optionally, a document including text that matches a the query terms with possible intermediate words may be considered a hit.
This is useful when the query terms need not occur consecutively in the document, such as in the case of ``left amygdala''  which should match ``left and right amygdala'' or ``left basolateral amygdala'' since they refer to left amygdala.
All punctuation marks within the document are ignored. 
 
Articles are represented as:
$<t,a,s_{s}>$ where \vart{t}{T}, \vart{a}{Text}, and \setvar{s}{S}.

The query expansion is performed as follows:

\begin{itemize}

\item{Query Component Processing:} {The query terms are searched in the title and the abstract of an article.
The query terms are converted to regular expressions and expanded with the acronyms and synonyms.
}

\item{Acronym and Synonym Expansion:} {To capture the intention of the query, domain specific acronyms and synonyms are used to expand the query terms. 
For example, \texttt{left amygdala} is expanded to ``\texttt{left} (\texttt{amygdala}$\mid$ \texttt{amygdalar}$\mid$\texttt{amygdalae})''}.

\item{Window of matching terms:}{
Queries containing multiple terms match sentences with up to six  words between the query words. 
The regular expressions are constructed to accept six intermediate words (a number that was determined through experimentation).
For example, the query term \texttt{left amygdala} becomes ``\texttt{left ((\textbackslash w)* )\{0,6\}\texttt{amygdala}''}}.

\item{Word Boundary Restrictions:} {Regular expressions impose word boundaries on query terms so that  a query term like ``\texttt{man}'' only matches ``\texttt{man}'' and not words that include it as a substring (performance, manner, manipulation, and woman).
The query term ``\texttt{man}'' is transformed to the regular expression ``\textbackslash W(\texttt{man})\textbackslash W''.}
\end{itemize}

Each query term is transformed into a regular expression such that the resultant set consists of all the matching documents.
The results of an inquiry are presented as a table that displays the frequencies of the documents for each cross-query. 
This enables viewing the relative frequencies of the combinations of the queries within an inquiry in a juxtaposed manner. 
The frequency of articles according the their year of publication are shown as a histogram. 
For a specific year, matching articles are shown with their \pmid, \doi, publication year, article title, and the sentences in the article that match the query. 
The user may view a specific article on \pubmed\ by clicking on the \pmid.

\section{Implementation Details}
\label{sec:implementation}

A prototype, NeuroBoun, of the proposed method is implemented as a web application using Python\footnote{ can be accessed at \textit{http://soslab.cmpe.boun.edu.tr/neuroboun} and the software is available on a git repository. The data is available via \pubmed.}
Documents were retrieved from PubMed via Entrez API \footnote{The Entrez API is available at: \url{https://eutils.ncbi.nlm.nih.gov/entrez/eutils/}. For ease of replicability, we provide a script to fetch the documents that may be found at the code repository.}.
A subset of the articles were fetched that contain the term ``amygdala'' (totaling 38,998 as of 12.10.2019).
The documents are represented with their PMID (document identifiers provided by \pubmed),  their titles and their abstracts which are stored as a list of sentences. 
Sentences are the main units of documents that are used to seek elements of interest. 
The NLTK library \cite{nltk} is used to parse the abstracts to yield sentences. 
PostgreSQL \cite{psql} was used to store the processed documents. 
A demo of the system is accessible at \url{soslab.cmpe.boun.edu.tr/neuroboun}.

\section{Use case}
\label{sec:use_case}

The use of \neuroboun\ tool will be described with a use case. 
A user can search a subject along with several dimensions.
The cross-queries (as described in Section~\ref{sec:approach}) are searched within the articles and the frequencies and percentages that match the articles are returned. 
As such, the researcher is presented the body of relevant publications for a quick perusal or to perform detailed analysis such as to inspect the results of specific cross-queries, specific documents, or sentences.

\begin{figure*}
\begin{center}
\frame{\includegraphics[width=0.8\textwidth]{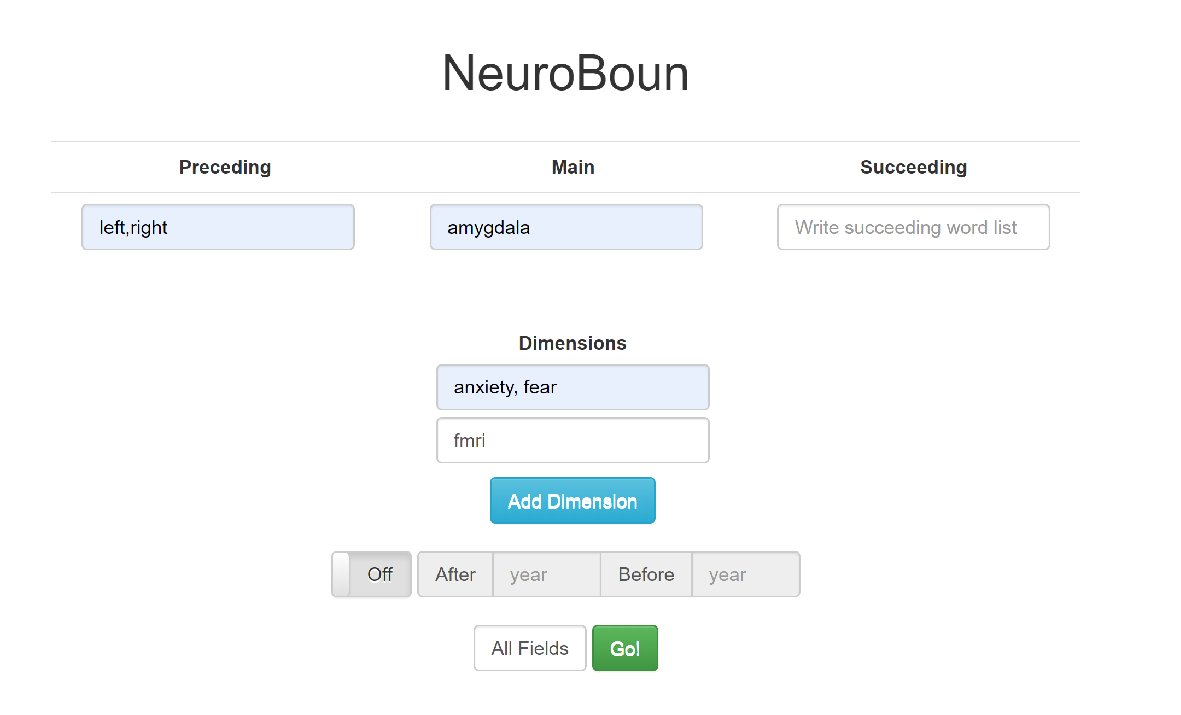}}
\caption{\label{fig:comparative-inquiry} An inquiry for inspecting  the lateralization of amygdala with two dimensions.}
\end{center}
\end{figure*}

To start with, \neuroboun\ offers several features to specify inquires as shown in Figure~\ref{fig:comparative-inquiry}.
The following describes the user interface elements that may be used to specify an inquiry:

\begin{description}[leftmargin=*]
\item[Preceding/succeeding word:] The main query may specify other terms that may precede or succeed the central term.
When preceding or succeeding words are specified a match within the document succeeds when the terms are in the same sentence and occur within a distance of zero to six words. 
In case of our example two preceding terms are specified (``left'' and ``right'').
The results will be presented separately for the documents that include ``amygdala'' and those that precede ``amygdala'' with the terms ``left'' and for ``right'' within 0-6 words.
For example, a document that has a sentence including  ``left basolateral amygdala'' will be considered a match.
In this manner the user will be able to observe the degree to which studies  consider the left and right of amygdala in research. 
\item[Dimension:] Users can specify additional dimensions of interest by specifying corresponding search terms.
Each dimension consists of one or more search terms and each term consist of one or more words.
These dimensions are searched within the articles to inspect the degree to which they co-occur in the same context.
The user is presented an overview table that summarizes the results of the cross-queries formed based on the dimensions. 
The results can be further inspected by selecting a desired cell within the summary table.
In our use case the dimensions were related to the brain regions (amygdala), moods (anxiety), and imaging technologies (fMRI).
\item[Time period:] The user may limit the articles according the the publication date (year) by specifying a year for before and/or after some given year. 
For example, the user may select ''After'' 1990 to explore the impact of the emergence of \fmri\ technology in research facilities.
\item[Field selection:] An inquiry can be performed by searching for the terms within the title and/or the abstract of the articles.
This is specified by selecting the desired option next to \textit{Go} button (default is ``All Fields''). 
\end{description}

In our case, we wanted to identify the articles related to amygdala to delineate the functional, neuroanatomical and neurophysiological differences between the left and the right amygdala in the mammalian brain. 
Additionally, we are interested in how development of new research tools including neuroimaging changed the depth and scope of research on amygdalar lateralization and whether there has been adequate explanation of the mechanism underlying the lateralization commensurate with the increasing rate of publications on this specific topic. 
Figure~\ref{fig:comparative-inquiry} shows this inquiry (\textit{lateralization of amygdala in the context of the moods of anxiety and fear}) using \neuroboun\ where the  main query is set to ``amygdala'' with the preceding words  ``left'' and ``right''.
Two dimensions are specified, one for moods (``anxiety'', ``fear'') and another for an imaging technology (``\fmri'').
No time limitation is specified.

\begin{figure*}
\begin{center}
\frame{\includegraphics[width=1\textwidth]{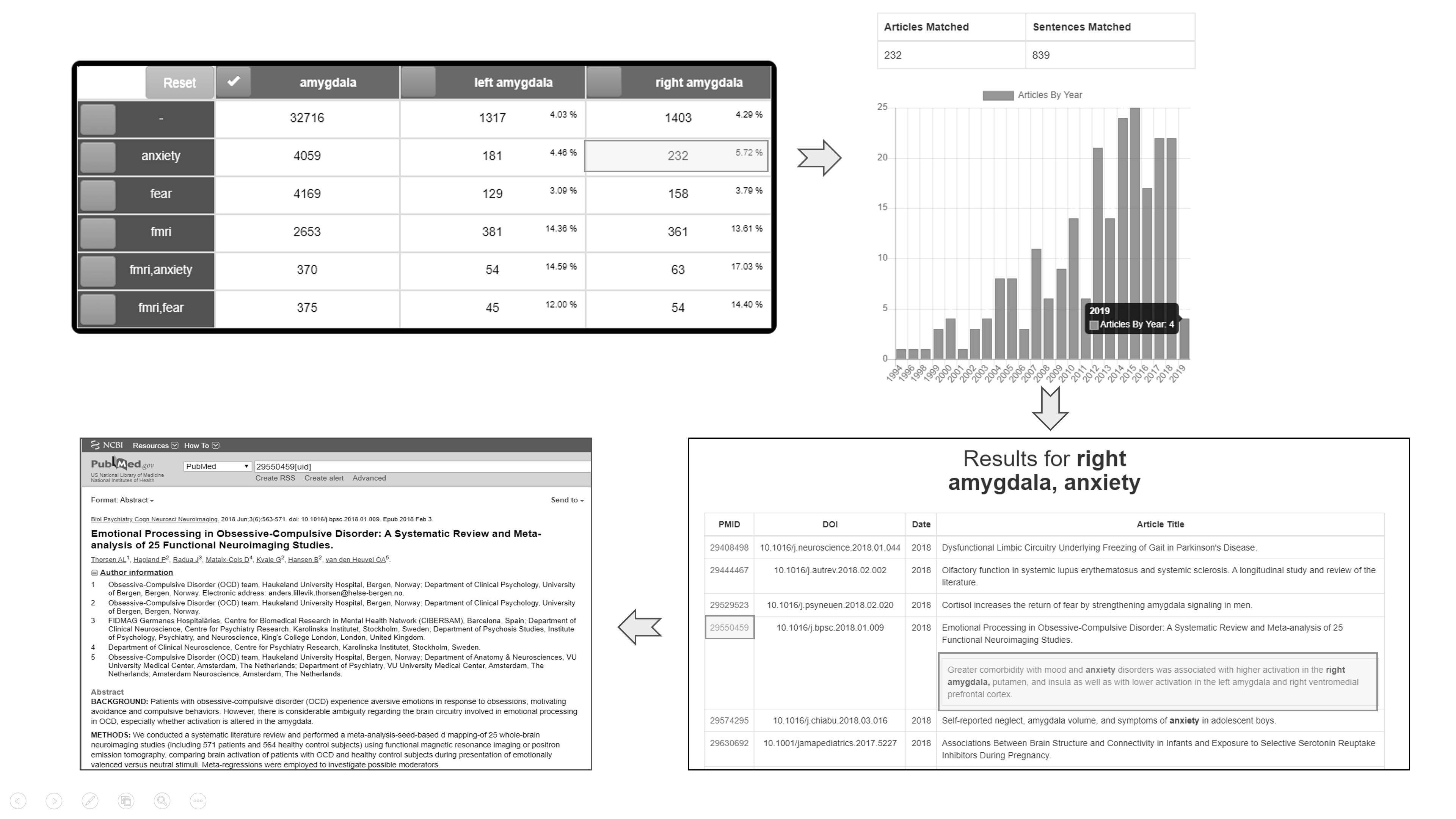}}
\caption{\label{fig:comparative-results} Search Results for main query of ``amygdala'' in comparison with ``left'' and ``right'' amygdala which is combined with dimensions of moods like anxiety, fear and \fmri}
\end{center}
\end{figure*}

Figure~\ref{fig:comparative-results} shows various screens resulting from this inquiry. The screen on the top left,  presents the frequency/percentages of each cross-query in a juxtaposed manner to reveal the degree to which they co-occur.
While examining the results, researchers can dive into the details of a specific cross-section by clicking in a cell (i.e. right amygdala and anxiety) which shows a trend graph of the number of articles published on a yearly basis is shown along with the total number of matching articles and sentences.
The user can further proceed by examining the documents that match the queries.
For each document the \doi, \pmid, title,  and a link to \pubmed\ article is shown.
More significantly, the relevant sentences may be seen which highlight the matching terms in the title and abstract of the articles. In this example, the query "right amygdala" matches with an article containing a sentence which has three words between "right" and "amygdala", namely "right or the left amygdala" (see  bottom right screen in Figure~\ref{fig:comparative-results}. 
Finally, one may continue their exploration by accessing specific articles in \pubmed\  by clicking on the id of an article.

\section{Discussion and Future Work}
\label{sec:discussion}

The main aim of this work is to provide researchers with a tool to investigate a subject of interest within a body of work through inquiries that embody a set of queries. 
For this purpose we developed a model for performing and aggregating inter-related queries.
A prototype of this model was implemented that presents the results in a juxtaposition manner to facilitate the comparison of the relative density of work regarding the aspects of the inquiry. 
It also, supports delving into any of the sub-contexts to  view the work over years and the specific articles that highlight the matching terms.

The current version of the system performs queries over the titles, abstracts, and keywords of articles from \pubmed. 
In recent years, there is a significant increase in open access articles which provides the full text of documents.
This exciting development presents many opportunities to improve our results. 
Per feedback from researchers (who used our tool), we think that the locations where the query terms match the document (\ie\ \textit{Future Work} and \textit{Results}) may provide critical clues for researchers regarding the significance of the papers or their findings.
Another issue of relevance is the assessment of the  qualitative contribution of an article, such as the introduction of a new method or an explanation. 

We plan to improve inquiries with the use of more advanced NLP methods and semantic web techniques to benefit from the standardized terms and the representation of domain knowledge to capture relationships within documents and to perform semantic processing. 

While the implementation of \neuroboun\ tool supports inquiry in the neuroscience domain, the design of the system is domain-independent with its own document model.
Thus, it can be adapted for alternative domains of interest.

Finally, we intend to publish a paper dedicated to a critical assessment of amygdalar asymmetry that uses the facilities provided by \neuroboun.

\section{Conclusion}
\label{sec:conclusions}

We have proposed an approach for inquiry over a published body of work and implemented a prototype that embodies its infrastructure.
The beneficial insights provided by this inquiry-based approach is demonstrated with the the prototype we developed that uses \pubmed. 
Although several approaches that utilize \pubmed\ results for alternative presentation exist, ours provides a unified view of several enhanced queries. 

\neuroboun\ was developed based on our need to achieve a comprehensive, efficient and expandable search for the case that we briefly present here, namely amygdalar lateralization. 
Although, its development originated from a specific context, it is designed to serve any body of knowledge.
We are encouraged by the results which for the foundation of work we intend to pursue in several directions, among the most significant are using domain-specific ontologies to semantically annotate and retrieve articles and extending the inquiries to search the full body of articles which are becoming increasingly accessible due to the support for open access publication.

\section*{Acknowledgements} The authors gratefully acknowledge  {\"Ozgur} Akyaz{\i} and {\c{S}}ahin Batmaz for their efforts in implementing the prototype and the members of SosLab and TABILAB of the Computer Engineering Department at Bogazici University for their valuable feedback throughout this work.

\bibliographystyle{natbib}
\bibliography{references.bib}

\end{document}